\documentclass[aps,prb,twocolumn,preprintnumbers,amsmath,amssymb]{revtex4}
\usepackage{graphicx}
\usepackage{float}
\usepackage{xcolor}
\begin{document}
\title{Magnetoresistance in the helical itinerant magnets MnSi and Mn$_{1-x}$Co$_x$Si}

\author{A.~E.~Petrova}
\affiliation{P.~N.~Lebedev Physical Institute, Leninsky pr., 53, 119991 Moscow, Russia}
\author{S.~Yu.~Gavrilkin}
\affiliation{P.~N.~Lebedev Physical Institute, Leninsky pr., 53, 119991 Moscow, Russia}
\author{D. Menzel}
\affiliation{Institut f\"{u}r Physik der Kondensierten Materie, Technische Universit\"{a}t Braunschweig, D-38106 Braunschweig, Germany}
\author{V.~A.~Stepanov}
\affiliation{P.~N.~Lebedev Physical Institute, Leninsky pr., 53, 119991 Moscow, Russia}
\author{S.~M.~Stishov}
\email{stishovsm@lebedev.ru}
\affiliation{P. N. Lebedev Physical Institute, Leninsky pr., 53, 119991 Moscow, Russia}

\begin{abstract}
We studied the longitudinal and transverse magnetoresistance of helical magnets, MnSi and Mn$_{1-x}$Co$_x$Si, at temperatures between 1.8 and 100~K and in magnetic fields up to 9 Tesla. All substances exhibited negative longitudinal and transverse magnetoresistance at temperatures above 4~K, which is most likely related to the suppression of spin fluctuations by the magnetic field. Note that in contrast to our finding, the longitudinal magnetoresistance of ferromagnetic metals was found to be positive. The unique positive and anisotropic magnetoresistance of pure MnSi at low temperatures (1.8 and 4~K) in the induced ferromagnetic phase shows effective suppression of fluctuations by the magnetic field. The significant difference in behavior between pure MnSi and doped MnSi lies in the specifics of the latter material, which forms a sort of helical fluctuation cloud and reveals quantum critical properties at low temperatures. The observed isotropic magnetoresistance in MnSi and Mn$_{1-x}$Co$_x$Si at higher temperatures can tentatively be attributed to the shortening of the mean free path of electrical carriers due to scattering on magnetic fluctuations and impurities, which results in a suppression of Lorentz force effects.

\end{abstract}

\maketitle

\section{Introduction}
MnSi, a model helical magnet, crystallizes in the B20 structural type. Its space group P2$_1$3 does not contain a center of symmetry, which allows for the existence of a helical (chiral) magnetic structure~\cite{stish, stish2}.
The temperature of the magnetic phase transition in MnSi is approximately 29~K at atmospheric pressure. A magnetic field destroys the helical magnetic spiral structure while simultaneously forming conical and skyrmionic spin configurations~\cite{stish2,Ishi,Bauer}. It is important to note that the violation of spatial symmetry in the B20 structural type results in the existence of special points, including Weyl points, in the electronic and phonon spectra~\cite{stish3}. This situation led to the discovery of negative longitudinal magnetoresistance of topological origin in the B20 compound CoSi~\cite{Sch, Bal}. This feature is probably not expected in MnSi. In simple metals, the transverse magnetoresistance (TMR) is positive in low magnetic fields, resulting from the action of the Lorentz force, which bends electron trajectories. This phenomenon is described by the simple expression~\cite{Abr, Abr2},
$\Delta \rho/\rho\sim (L/r_{L})^{2}\sim H^{2}$, where $\rho$ is the electrical resistance,  $L$ is the mean free path of the carriers, $r_{L}$ is the Larmor radius, and $H$ is the magnetic field.

Meanwhile, the transverse magnetoresistance of MnSi was described as negative in Ref.~\cite{kad,min,neu} in low magnetic fields as is observed in metallic ferromagnets~\cite{Blat,Cam}. In ferromagnets, conduction electrons (s-electrons) scatter on localized fluctuating magnetic moments (d-electrons). A magnetic field can align these moments, reducing scattering and leading to negative magnetoresistance (MR). The situation is unclear with respect to longitudinal magnetoresistance (LMR). A number of classical papers have shown that LMR is positive in typical ferromagnetic metals such as nickel and iron~\cite{heap,pot}, though fluctuating magnetic moments should also influence MR in the longitudinal regime. Therefore, investigating the behavior of LMR in helical magnets, such as MnSi, where fluctuations of the magnetic moment are strong, is of great interest.

\section{Experimental} 

\begin{figure}[htb]
\includegraphics[width=80mm]{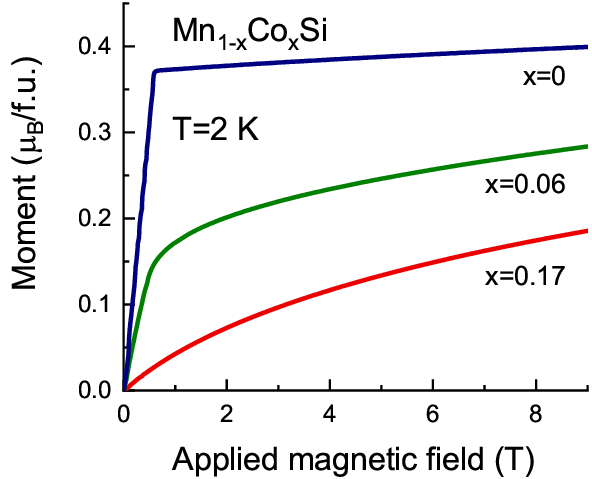}
\caption{\label{fig1} Magnetization curves for MnSi and Mn$_{1-x}$Co$_x$Si.} 
\end{figure}

\begin{figure}[htb]
	\includegraphics[width=80mm]{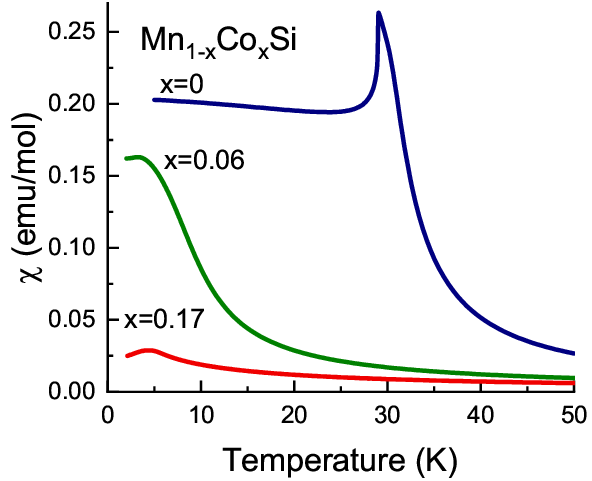}
	\caption{\label{fig2} Magnetic susceptibility of MnSi and Mn$_{1-x}$Co$_x$Si as a function of the temperature at 0.03~T.}
\end{figure}

This paper focuses on both the transverse and longitudinal magnetoresistance study of MnSi and Mn$_{1-x}$Co$_x$Si at $x=0.06$ and $0.17$. One reason for including the mixed compound Mn$_{1-x}$Co$_x$Si in the study is the expectation of effects of stronger magnetic moment fluctuations than those in MnSi, especially at low temperatures in the quantum critical region~\cite{Pet, Pet2, stish4}. The physical characteristics of the studied MnSi and Mn$_{1-x}$Co$_x$Si samples are described in Refs.~\cite{stish5, Pet, Pet2}. The magnetic properties of the samples are illustrated in Figs.~\ref{fig1} and \ref{fig2}. 

Magnetoresistance was measured using the standard four-terminal scheme with gold wires bonded to the sample by silver paste as electrical contacts in two orientations, with the current either parallel ($I\| H$) or perpendicular ($I\bot H$) to the magnetic field. During the measurements, the sample was cooled stepwise from 300~K to 1.8~K. At each step, the magnetic field was increased from 0 to 9~T, and the resistance was measured in the $I\bot H$ configuration. Then, the measurement continued in the $I\|H$ configuration as the magnetic field decreased from 9~T to 0. 
The experimental results on MR of MnSi and Mn$_{1-x}$Co$_x$Si are displayed in Figs.~\ref{fig3}, \ref{fig4}, and \ref{fig5}. 

\section{Discussion}
As can be seen in Figs.~\ref{fig1} and \ref{fig2}, only pure MnSi exhibits a distinct magnetic phase transition. At low temperatures, MnSi undergoes a phase transition from a spin helix into a conical spin state when an external magnetic field is increased. Finally, it ends up in a field-polarized phase, where the spins are aligned parallel to each other. The sharp slope change in the MR curves at 1.8 and 4~K occurs at the border between the conical and field-induced phases of MnSi ~\cite{stish2,Ishi,Bauer}. At higher temperatures, this border smears out, reflecting the response of the magnetic moment of MnSi to the magnetic field at elevated temperatures~\cite{min}. Nevertheless, the transition from the conical phase to the field-induced phase can still be observed in the magnetoresistance curves at 7~K and 10~K, as indicated by the arrows (Fig.~\ref{fig4}a).

\begin{figure}[htb]
	\includegraphics[width=60mm]{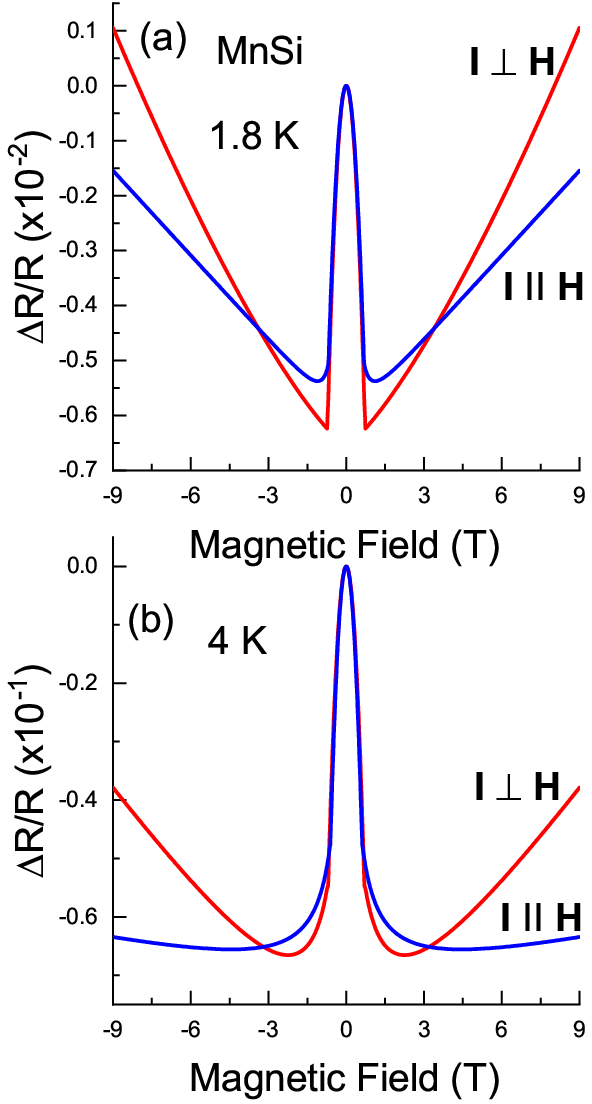}
	\caption{\label{fig3} Longitudinal and transverse magnetoresistance of MnSi as a function of the magnetic field at (a) 1.8~K and (b) 4~K.}
\end{figure}

As can be seen in Fig.~\ref{fig3} the magnetoresistivity curves at 1.8~K and 4~K corresponding to pure MnSi differ drastically from the rest of the data depicted in Figs.~\ref{fig4} and \ref{fig5}. Figs.~\ref{fig1} and \ref{fig2} provide some insight into the situation by illustrating the basic features of the MnSi phase diagram and its magnetic properties. As demonstrated in Fig.~\ref{fig3}, the helical and conical phases of MnSi show a negative response to low applied magnetic fields at 1.8~K and 4~K, which is obviously a result of suppressing weak magnetic fluctuations. However, in the field-induced phase at low temperature, the MR of MnSi becomes positive, indicating the action of the Lorentz force, which normally leads to a "quadratic" field dependence ($\Delta \rho/\rho\sim H^{n}$, although the exponent $n$ rarely reaches the canonical value of two). In our case, $n\approx1.5$ for the transverse branch of MR at T=1.8~K. 

\begin{figure}[htb]
	\includegraphics[width=60mm]{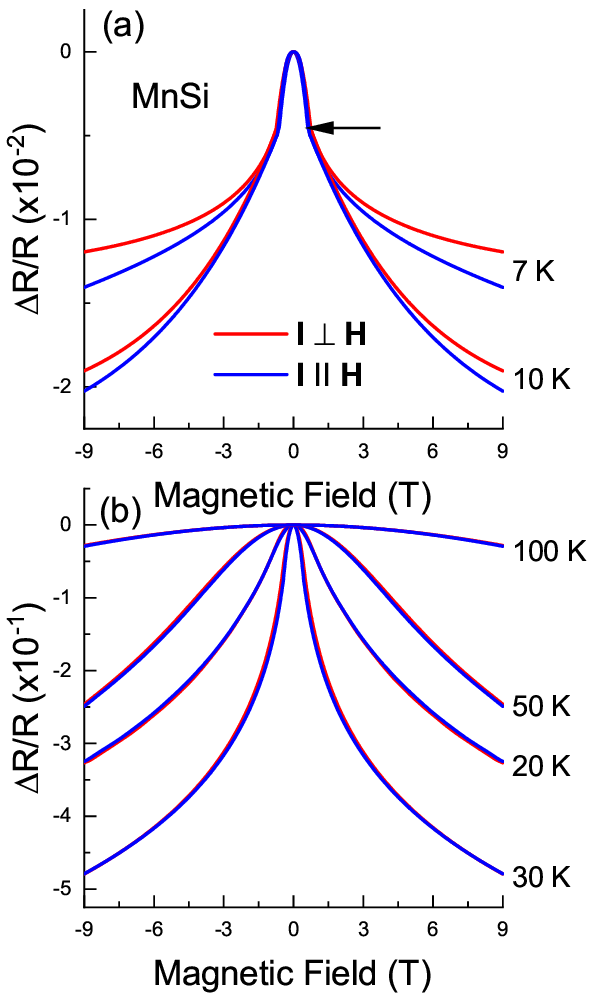}
	\caption{\label{fig4}  Magnetoresistance of MnSi as a function of the magnetic field at 7~K and 10~K (a) and at 20~K, 30~K, 50~K, 100~K (b). At temperatures above 20~K the longitudinal and transverse magnetoresistance curves can't be distinguished. Arrow indicates border between conical and field polarized phases.}
\end{figure}
     
As is seen in Fig.~\ref{fig3} the MR data of MnSi at 1.8~K and 4~K clearly show anisotropic character, for magnetic fields above 0.5~T. Significantly, the anisotropy of MR is not observed in the helical phase, probably due to the masking effect of strong magnetic fluctuations. Suppressing magnetic fluctuations with a magnetic field extends the path of charge carriers due to the Lorentz force. Thus, a positive and anisotropic field dependence of MR is observed, which changes rapidly with temperature. As the temperature increases, the MR in MnSi becomes negative and practically isotropic, as can be seen in Fig.~\ref{fig4}, which shows the MR data for Mn$_{1-x}$Co$_x$Si at $x=0.06$ and $0.17$. 

\begin{figure}[htb]
	\includegraphics[width=70mm]{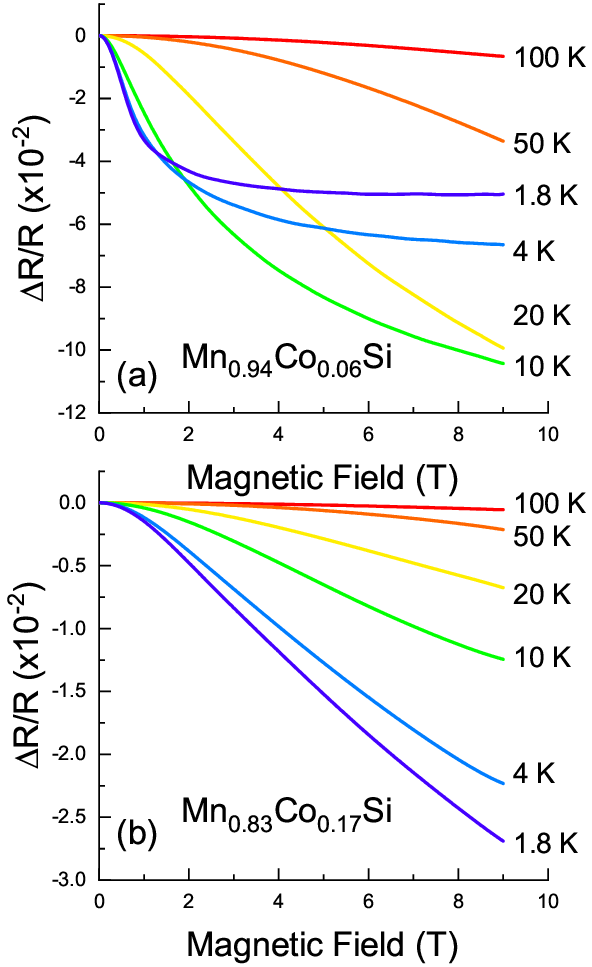}
	\caption{\label{fig5}  Magnetoresistance of (a) Mn$_{0.94}$Co$_{0.06}$Si and (b) Mn$_{0.83}$Co$_{0.17}$Si as a function of the magnetic field. The longitudinal and transverse magnetoresistance can not be distinguished in the present experiment.}
\end{figure}

\begin{figure}[htb]
\includegraphics[width=85mm]{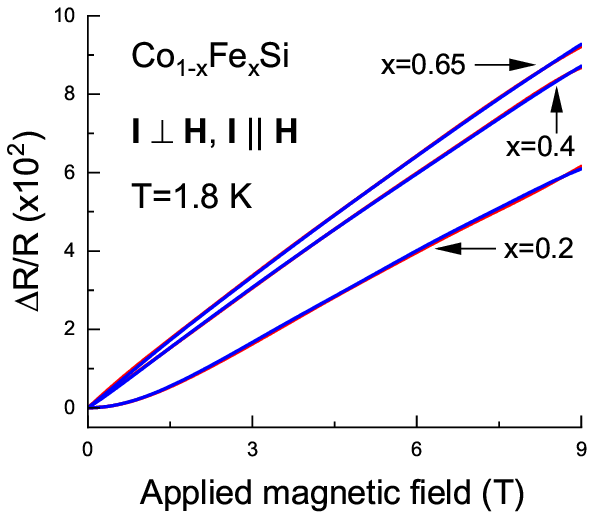}
\caption{\label{fig6} Longitudinal and transverse magnetoresistance of Co$_{1-x}$Fe$_{x}$Si as a function of the magnetic field at 1.8~K~\cite{Pet4}.}
\end{figure}

Remarkably, as can be seen in Figs.~\ref{fig4} and \ref{fig5}, the magnetoresistance significantly decreases with the amount of Co impurity.  Indeed, the magnetoresistance of Mn$_{0.83}$Co$_{0.17}$Si at a comparable temperature and magnetic field is only about one-tenth that of MnSi. 
Given the isotropic properties of the substances under study, one might suppose that the Lorenz force plays a minor role in magnetoresistance. Therefore, scattering of carriers on magnetic fluctuations and impurities is the main cause of electrical resistance in these materials. It is important to note that the observed extraordinary isotropic behavior of MR in the studied materials is not directly related to its negative sign. An impressive example of isotropic behavior exhibiting positive magnetoresistance is the related compound Co$_{1-x}$Fe$_x$Si (see Fig.~\ref{fig6}). In this case, Lorenz force effects are probably suppressed because the mean free path of the carriers is too short due to scattering on impurities, making all directions in a sample equivalent with respect to the magnetic field.

It is important to emphasize the significant difference in the behavior of MR in pure and doped MnSi related to the specific properties of the latter material. As mentioned in~\cite{Pet2}, doping MnSi with Co completely destroys the helical phase transition but preserves the helical fluctuation region, which is normally located slightly above the phase transition temperature. However, at Co doping this area, which spreads from approximately 5~K to 0~K, forms a sort of helical fluctuation cloud that reveals quantum critical properties: $C_p/T\rightarrow \infty$ at $T \rightarrow 0$, which is not easily suppressed by a magnetic field.

\section{Conclusion}

We studied the  magnetoresistance of helical magnets MnSi and Mn$_{1-x}$Co$_x$Si at temperatures between 1.8 and 100~K and in magnetic fields up to 9 Tesla. 
The magnetoresistivity of pure MnSi at 1.8~K and 4~K (Fig.~\ref{fig3}), which is positive and anisotropic, differs drastically from the rest of the data depicted in Figs.~\ref{fig4} and \ref{fig5}. At the same time, all substances exhibit negative longitudinal and transverse magnetoresistance at temperatures above 4~K, which is most likely related to the suppression of magnetic fluctuations by the magnetic field. Note that, in contrast to our finding, the longitudinal magnetoresistance in ferromagnetic metals was found to be positive~\cite{heap, pot}.
The exceptional behavior of the magnetoresistance of pure MnSi at low temperatures (1.8~K and 4~K) is probably related to annihilating weak magnetic fluctuations by low magnetic fields (Fig.~\ref{fig3}). 
The magnetoresistance significantly decreases with the amount of Co impurity. Indeed, the magnetoresistance of Mn$_{0.83}$Co$_{0.17}$Si at a comparable temperature and magnetic field is only about one-tenth that of MnSi, as can be seen in Figs.~\ref{fig4} and \ref{fig5}. Consequently, the MR of Mn$_{1-x}$Co$_x$Si at high temperatures is mostly defined by the influence of the magnetic field on the impurity and phonon contributions to electrical resistance. Nevertheless, a slight negative slope of $\Delta \rho/\rho (H)$ is still observed even at 100~K, indicating that magnetic fluctuations survive at high temperatures despite the disorder caused by the Co impurity (Fig.~\ref{fig5}). 
The observed isotropic of magnetoresistance in MnSi and Mn$_{1-x}$Co$_x$Si can tentatively be attributed to the shortening of the mean free path of electrical carriers due to scattering on magnetic fluctuations and impurities. This results in the suppression of Lorentz force effects. The magnetic fluctuations in the quantum regime at low temperatures of Mn$_{1-x}$Co$_x$Si appear to be much stronger than in MnSi, as we initially guessed. Therefore, we could not observe the canonical MR behavior of Mn$_{1-x}$Co$_x$Si, which is positive and anisotropic, as shown in Fig.~\ref{fig3} for MnSi.

Finally:
1. The longitudinal and transverse magnetoresistances of the helical magnet MnSi were measured at temperatures between 1.8 and 100~K and in magnetic fields up to 9 Tesla.\\
2.  The longitudinal and transverse magnetoresistance of Mn$_{1-x}$Co$_x$Si at $x=0.06$ and $0.17$ were measured within the same temperature and magnetic field range.\\
3. The magnetoresistivity of pure MnSi at 1.8~K and 4~K (Fig.~\ref{fig3}) differs drastically from the rest of the data depicted in Figs.~\ref{fig4} and \ref{fig5}, being positive and anisotropic.  
4. The longitudinal and transverse magnetoresistance of all the substances studied appeared to be negative at temperatures higher than 4 K due to the suppression of magnetic fluctuations by magnetic fields. Some of our results contrast with data on ferromagnetic metals, whose longitudinal magnetostriction appears positive.\\
5. Almost all measured magnetoresistance data are highly isotropic, meaning their values do not depend on the relative positions of the electrical current and magnetic field directions, with the exception of pure MnSi at very low temperatures. The observed isotropic magnetoresistance in MnSi and Mn$_{1-x}$Co$_x$Si may be due to the shortening of the mean free path of electrical carriers caused by scattering from magnetic fluctuations and impurities.\\
6. The magnetoresistance decreases significantly with the amount of Co impurity. Indeed, the magnetoresistance of Mn$_{0.83}$Co$_{0.17}$ at a comparable temperature and magnetic field is only about one-tenth that of MnSi, clearly indicating the suppression of magnetic moments and associated magnetic fluctuations.\\
7. A significant difference in the behavior of MR in pure MnSi and doped MnSi is related to the specifics of the latter material. Doping MnSi with Co completely destroys the helical phase transition but preserves the helical fluctuation area. This area forms a helical fluctuation cloud at low temperatures, revealing quantum critical properties~\cite{Pet,Pet2,stish4}.

\end{document}